\pdfoutput=1
\documentclass[fleqn,usenatbib]{mnras}

\usepackage{newtxtext,newtxmath}

\usepackage[T1]{fontenc}

\usepackage[dvipsnames]{xcolor}
\usepackage[normalem]{ulem}
\usepackage{bm}
\usepackage[utf8]{inputenc}
\usepackage{pgfplots}
\usepackage{tikz}
\usepgfplotslibrary{groupplots}
\pgfplotsset{compat=1.14}
\usepackage{capt-of}
\usepackage{graphicx}

\usepackage{amssymb}

\usepackage{mathrsfs}
\usepackage{amsmath}
\usepackage{hyperref}
\usepackage{lineno}
\usepackage{subfigure}
\usepackage{widetext}

\title[Small-scale radio jets and TDEs]{Small-scale radio jets and tidal disruption events: A theory of high-luminosity compact symmetric objects}

\author[A.G. Sullivan et al.]{
Andrew G. Sullivan$^{1}$,\thanks{E-mail: ags2198@stanford.edu} 
Roger D. Blandford$^{1}$,
Mitchell C. Begelman$^{2,3}$,
Mark  Birkinshaw$^{4}$,
\newauthor
Anthony C.S. Readhead$^5$ 
\\
$^{1}$Kavli Institute for Particle Astrophysics and Cosmology, Department of Physics, Stanford University, Stanford, CA 94305, USA\\
$^{2}$JILA, University of Colorado and National Institute of Standards and Technology, 440 UCB, Boulder, CO 80309-0440, USA\\
$^{3}$Department of Astrophysical and Planetary Sciences, University of Colorado, 391 UCB, Boulder, CO 80309-0391, USA\\
$^4$School of Physics, H.H. Wills Physics Laboratory, University of Bristol, Tyndall Avenue, Bristol BS8 1TL, UK\\
$^5$Owens Valley Radio Observatory, California Institute of Technology,  Pasadena, CA 91125, USA
}

\date{Accepted 2024 January 24. Received 2024 January 23; in original form 2023 December 7}

\pubyear{2024}

\begin{document}
\label{firstpage}
\pagerange{\pageref{firstpage}--\pageref{lastpage}}
\maketitle

\begin{abstract}
Double lobe radio sources associated with active galactic nuclei represent one of the longest studied groups in radio astronomy. A particular sub-group of double radio sources comprises the compact symmetric objects (CSOs). CSOs are distinguished by their prominent double structure and sub-kpc total size.  It has been argued that the vast majority of  high-luminosity CSOs (CSO 2s) represent a distinct class of active galactic nuclei with its own morphological structure and life-cycle. In this work, we present theoretical considerations regarding CSO 2s. We develop a semi-analytic evolutionary model, inspired by the results of large-scale numerical simulations of relativistic jets, that reproduces the features of the radio source population. We show that CSO 2s may be generated by finite energy injections and propose stellar tidal disruption events as a possible cause. We find that tidal disruption events of giant branch stars with masses $\gtrsim1$ M$_\odot$ can fuel these sources and discuss possible approaches to confirming this hypothesis. We predict that if the tidal disruption scenario holds, CSO 2s with sizes less than 400 pc should outnumber larger sources by more than a factor of $10$. Our results motivate future numerical studies to determine whether the scenarios we consider for fueling and source evolution can explain the observed radio morphologies.
Multiwavelength observational campaigns directed at these sources will also provide critical insight into the origins of these objects, their environments, and their lifespans.
\end{abstract}

\section{Introduction} \label{sec:intro}

Compact Symmetric Objects (CSOs) represent a class of compact ($<1$ kpc) bright double radio sources, in which the radio lobes straddle the active galactic nucleus (AGN) \citep{1994ApJ...432L..87W,1996ApJ...460..612R}. There are two fundamentally different types of CSOs: an edge-dimmed, low-luminosity class (CSO 1) and and edge-brightened high-luminosity class (CSO 2) \citep{2016MNRAS.459..820T}. This paper focuses on CSO 2s. 

While some CSO 2s may be early stage Fanaroff-Riley-type (FR) double radio sources \citep{1974MNRAS.167P..31F}, it has been suggested that CSO 2s represent a separate class of AGN with different driving mechanisms and morphological history \citep{1994cers.conf...17R,1996ApJ...460..612R,PaperIII}. Such distinctions are manifest in the sizes and ages of CSO 2s relative to their FR counterparts \citep{PaperII}. Namely, specific CSO population analyses \citep{PaperI, PaperII} have shown that there exists a sharp size cutoff above 500 pc.  The size cutoff and their rapid hot spot separation speeds $\sim0.3 c$ \citep{1998A&A...337...69O, 1999NewAR..43..669O} suggest that CSO 2s generally have short $\lesssim 10^4$ yr lifespans.

Based on  morphology, the CSO 2 class may be subdivided into three sub-classes: CSO 2.0s, CSO 2.1s, and CSO 2.2s \citep{PaperIII}. \cite{PaperIII} argue that these three classes may represent the temporal evolution of CSO 2 objects, with CSO 2.0s as early life sources, CSO 2.1s as mid-life sources, and CSO 2.2s as late life sources. Because of the short lifetimes and low total energies of CSO 2s relative to FR Is and FR IIs, \cite{PaperIII} propose that tidal disruption events (TDEs) may be associated with CSO 2s, as had first been suggested by \citet{1994cers.conf...17R}.  Beyond the qualitative discussion of \cite{PaperIII}, a coherent theoretical model of the evolution of CSO 2s has yet to be proposed in light of this recent work.

Continuing the work of \cite{PaperIII}, we develop a theory of CSO 2 evolution, and discuss the open questions that observational campaigns can answer. We model the propagation of CSO 2 jets into their environment and consider their fueling mechanism, particularly the TDE scenario, as well as host galaxy environments. In Sec. \ref{sec:summary}, we summarize the recent work on CSO 2s motivating this paper. In Sec. \ref{sec:CSOevolution}, we describe the possible evolutionary model for CSO 2s and the features of CSO 2 morphological evolution. In Sec. \ref{sec:Disappearance}, we discuss the absence of CSO 2s larger than $500$ pc and its potential relationship to host galaxy environments. In Sec. \ref{sec:fueling}, we assess TDEs as a candidate fueling mechanism for CSO 2s, before concluding in Sec. \ref{sec:conclusion}.

\section{Summary of Compact Symmetric Object Observations}
\label{sec:summary}
CSO 2s are distinguished from other jetted AGN by their double lobe morphology, small sizes, and lack of features indicative of relativistic beaming towards the observer. Historically, double lobe sources with sizes less than 1 kpc
have been classified as CSOs, distinguishing them from medium symmetric objects (MSOs) with sizes $1-20$ kpc, and large symmetric objects (LSOs) with sizes greater than 20 kpc \citep{1995A&A...302..317F, 1996ApJ...460..612R}. Recently, \cite{PaperI, PaperII, PaperIII} presented an updated catalog of CSO 2s after introducing additional classification criteria to explicitly exclude blazars that mimic the compact double structure: 1) that source variability be slower than the light travel time across the emitting region and 2) that apparent superluminal motion in the source not exceed $2.5c$. 

The updated CSO catalog includes 79 \textit{bona fide} CSO 2s, 54 of which have measured spectroscopic redshifts \citep{PaperI}.  The distribution of sizes of the \textit{bona fide} CSO 2s cuts off sharply, well below 1 kpc, and signals a discontinuity in the size distribution of doubles around the CSO-MSO boundary. While CSO 2s have often been considered the early life versions of larger FR objects, they make up a much larger fraction of radio surveys than expected from a uniform size distribution \citep{PaperII}. In fact, fewer  than $5 \%$ of CSO 2s should evolve into the larger double radio sources to respect the population statistics of MSOs and LSOs \citep{PaperII}.

\begin{figure*}
    \centering
    \includegraphics[width=0.8\linewidth]{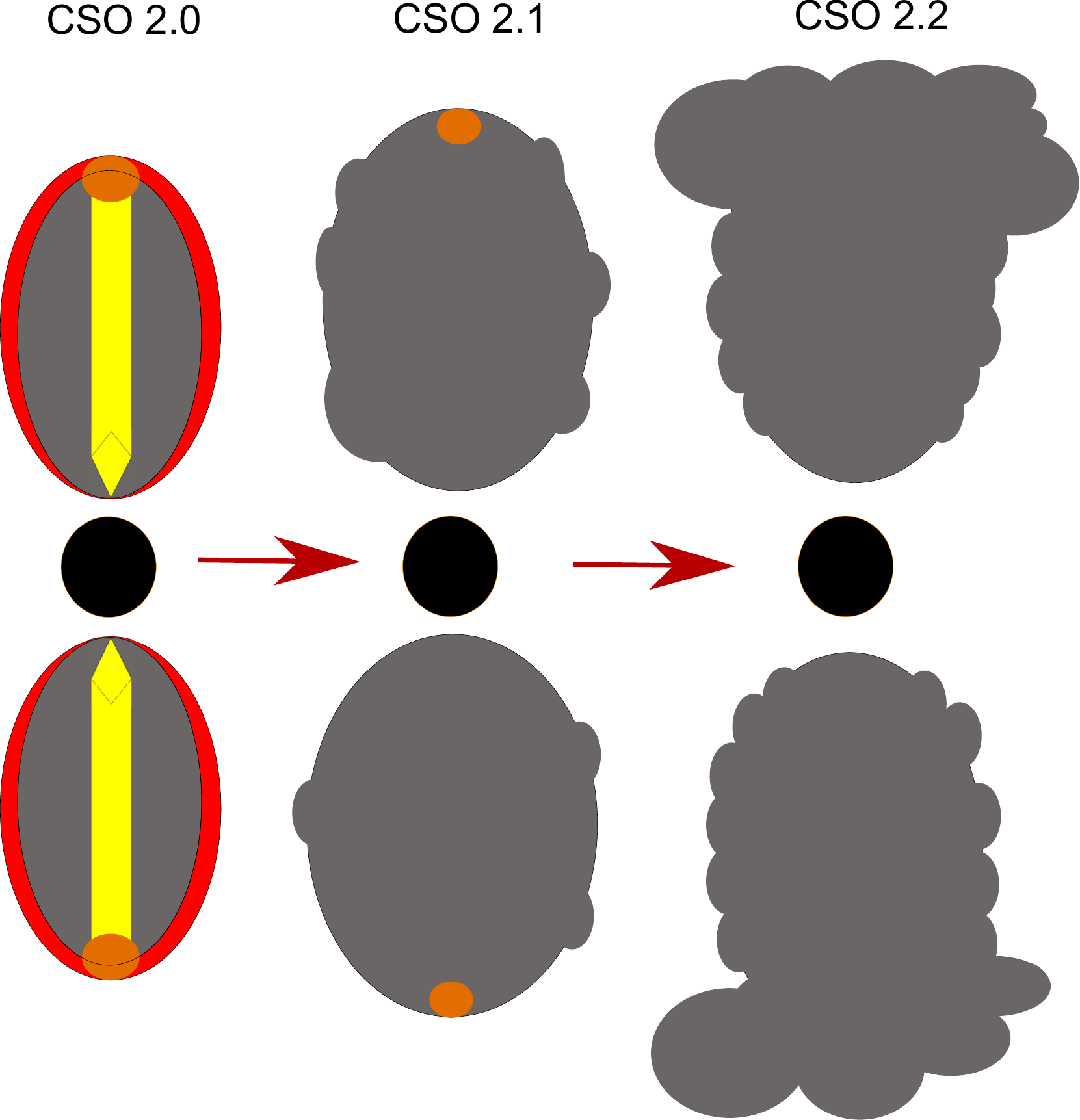}
    \caption{A diagram of the proposed evolutionary sequence of CSO 2s based on their radio morphology. The core is represented by the black circle. The radio lobes, which increase in size through the sequence, contain narrow jets represented in yellow and hot spots when young at early times. These fade while the lobes become more turbulent. See Fig. \ref{plt:CSO2.0_drawing} for a more detailed look at the CSO 2.0 case.}
\label{plt:Model_drawing}
\end{figure*}

The updated \cite{PaperI} \textit{bona fide} CSO catalog bifurcates into 11 CSO 1s and 43 CSO 2s with spectroscopic redshift \citep{PaperIII}. Lower power and smaller than CSO 2s, the CSO 1s have few common features across the sample. The CSO 2s are roughly evenly distributed into 2.0s, 2.1s, and 2.2s \citep{PaperIII}. CSO 2.0s have narrow lobes and prominent terminal hot spots in contrast to CSO 2.2s with wider lobes and no hot spots. CSO 2.1s represent a hybrid between CSO 2.0s and CSO 2.2s, containing features of both classes. The classified CSO 2.0s are by and large brighter and smaller than CSO 2.2s, with CSO 2.1s intermediate between the two. CSO 2.0s possess large hot spot separation speeds $\sim0.3c$, while those of CSO 2.2s are negligible. The standing hypothesis is that the CSO 2.0-2.1-2.2 paradigm maps the early, middle, and late stages of CSO 2 evolution \citep{PaperIII}. We sketch the CSO evolutionary sequence in Fig.~\ref{plt:Model_drawing}.

The CSO 2 luminosity-size distribution  (Fig.~1 in \citealt{PaperIII}) broadly shows an increase in luminosity with size among CSO 2.0s. This observed increase is not intrinsic, and results from the angular size cutoff at 4 milliarcseconds as well as the CSO luminosity function \citep{PaperIII}. Classic FR-II models propose that luminosity $L$ should decrease with size $D$ as $L\propto D^{-1/2}$ \citep{1997ApJ...487L.135R, 1997MNRAS.292..723K}. Should this relationship persist from the smallest FR-IIs ($\sim 10$ kpc) down to the smallest CSO 2s ($\sim10$ pc), one would expect 10 pc-sized CSO 2s with luminosities a factor of $\sim30$ brighter than the brightest FR-IIs ($\sim10^{29}$ W/Hz). The absence of these sources in the known samples implies that the luminosities of double radio sources must generally increase with size, and by extension time, for at least some portion of their early evolution. Consequently, the luminosities of CSO 2s (including those which develop into MSOs and LSOs) likely increase initially but fade at later times.

\section{Evolutionary Model of Compact Symmetric Objects}
\label{sec:CSOevolution}
\subsection{Early life}
\label{sec:Earlylife}
\begin{figure}
    \centering
    \includegraphics[width=0.8\linewidth]{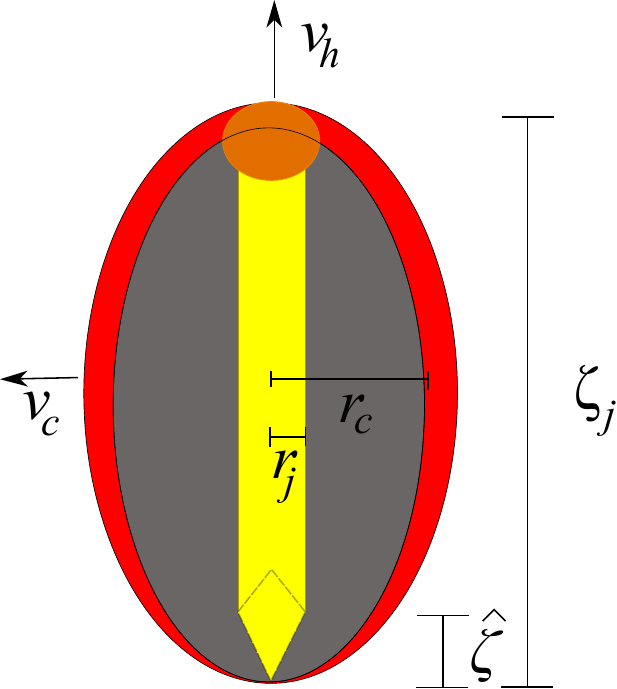}
    \caption{A detailed diagram of CSO 2.0s as modeled. The jet and inner cocoon radii are each labeled. We indicate the height of the recollimation shock. The red outer layer represents the shocked external medium.}
\label{plt:CSO2.0_drawing}
\end{figure}
Early life CSO 2s appear to evolve analogously to the larger and longer-lived classical FR-II objects. Their edge-brightened lobes with prominent hot spots are consistent with typical features of a relativistic jet propagating into the external medium. Classic jet models suppose that the jet collides with the ambient medium, forming a contact discontinuity between a forward and a reverse shock \citep[e.g.]{1984RvMP...56..255B, 1991MNRAS.250..581F, 1997MNRAS.286..215K, 2011ApJ...740..100B}. The material in the reverse shock becomes deflected upstream into a cocoon around the jet. The cocoon also produces synchrotron radiation, and accounts for dimmer radio emission upstream of the hot spots \citep{1984RvMP...56..255B, 1996ApJ...460..612R}.  The jet power $L_j$ determines how rapidly the jet pushes into the external medium. By analogy, CSO 2.0s should possess the aforementioned properties \citep{1996cyga.book..209B}. The radio hot spots of early life CSO 2s represent the working surface of the jet where particle heating and magnetic field amplification occur \citep{1984RvMP...56..255B}, while a cocoon forms around the narrow jet.

\subsubsection{Relativistic Jet Model} 
\label{subsubsec:ActiveJet}
Because early-life CSO 2s exhibit features so similar to those of classical FR-II sources, we apply the \cite{2011ApJ...740..100B} unmagnetized and gas pressure collimated jet model. The jet working surface, whose height we denote by $\zeta_j(t)$, advances with speed \citep{2011ApJ...740..100B}
\begin{equation}
\label{eq:headvelocity}
    v_h\approx\sqrt{\frac{L_j}{\rho_a A_j c}} ,
\end{equation}
where $\rho_a$ is the ambient density of the external medium, $A_j$ is the cross sectional area of the jet, and $c$ is the speed of light. 
The cocoon pressure $P_c$ collimates the initially conical jet, forming a shock, above which the now cylindrical jet has radius $r_j$ so that $A_j=\pi r_j^2$. The height of the collimation shock expands along the Mach cone as \citep{2011ApJ...740..100B}
\begin{equation}
\label{eq:collimation}
    \hat{\zeta}= \sqrt{\frac{L_j }{P_c \pi c}}.
\end{equation}
The collimation radius $r_j$ can be approximated as $r_j=1/2\theta_0 \hat{\zeta}$, where $\theta_0$ is the initial jet opening angle \citep{2011ApJ...740..100B}.
The cocoon pressure is
\begin{equation}
\label{eq:cocoonpressure}
    P_c=\frac{1}{3}\frac{E(t)}{\pi r_c^2 \zeta_j},
\end{equation}
where $E(t)$ is the energy deposited into the cocoon and $r_c$ is its radius. 
$P_c$ balances the external medium ram pressure, giving \citep{1989ApJ...345L..21B, 2011ApJ...740..100B}
\begin{equation}
\label{eq:cocoonvelocity}
v_c=\sqrt{\frac{P_c}{\rho_a}}\end{equation}
as the expansion rate of $r_c$.

Eq. \ref{eq:headvelocity}, \ref{eq:collimation}, \ref{eq:cocoonpressure}, and \ref{eq:cocoonvelocity} can be arranged into a system of differential equations for the height and cross-sectional areas of the cocoon and jet
\begin{subequations}
\label{eq:differentialequations}
    \begin{equation}
        \frac{d\zeta_j}{dt}=\sqrt{\frac{L_j}{\rho_a A_j c}},
    \end{equation}
    \begin{equation}
        \frac{dA_c}{dt}=2\sqrt{\frac{\pi E(t)} {3 \zeta_j \rho_a  c}},
    \end{equation}
    \begin{equation}
        A_j=\frac{3}{4}\theta_0^2 \frac{L_j}{E(t)c} A_c \zeta_j,
    \end{equation}
\end{subequations}
with $A_c=\pi r_c^2$ as the cocoon cross-sectional area. In a real source, $A_j$ may be wider if the hot spot wiggles around as in Cygnus A.
Note that $dA_c/d\zeta_j=\theta_0 \sqrt{\pi A_c}$ has solution $A_c=1/4\pi\theta_0^2 \zeta_j^2$ and allows eq. \ref{eq:differentialequations} to simplify to 
\begin{equation}
\label{eq:jetheight}
    \frac{d\zeta_j}{dt}= 4\sqrt{\frac{ E(t)}{3\pi\rho_a \theta_0^4 \zeta_j^3}}.
\end{equation}
In practice, this is an upper limit on the expansion speed because of the possible variation in $A_j$.

Because CSO luminosity increases with size, CSO 2.0 jet power may increase in kind. Therefore, we give the CSO jet power $L_j$ the phenomenological profile
\begin{equation}
\label{eq:jetpower}
    L_j(t)=\begin{cases}
        L_{max} \left(\frac{t}{t_0}\right)^{p_1} & t<t_0\\
        L_{max}   \left(\frac{t}{t_0}\right)^{-p_2} & t>t_0
    \end{cases}.
\end{equation}
Integrating eq. \ref{eq:jetpower} gives the energy deposited by the jets into the cocoon as a function of time $E(t)$.

Solving eq. \ref{eq:jetheight} with $E(t)$ and a prescription for the ambient density profile gives the height of the working surface as a function of time.
For $\rho_a=\rho_0 (\zeta_0/\zeta)^a$, the jet has a characteristic scale height $\zeta_s\equiv\left(\frac{L_{max}t_0^3}{\pi \theta_0^4 \rho_0 \zeta_0^a }\right)^{\frac{1}{5-a}}$. Illustrative parameters (e.g. $L_j=10^{44}$ erg s$^{-1}$, $t_0=2500$ yr, $\theta_0=10^{\circ}$, $a=1$, and $\rho_a=10^{-23}$ g cm$^{-3}$ at $\zeta_0=5$ pc) give $\zeta_s\approx190$ pc and $v_h\approx 0.15c$, which are consistent with the observations.

\subsubsection{Alternative Jet Models}
The model we have considered above is purely hydrodynamic \citep{1991MNRAS.250..581F, 1997MNRAS.286..215K, 2011ApJ...740..100B, 2018MNRAS.477.2128H}, although large-scale magnetically collimated jet models exist \citep{2002ApJ...572..445L, 2009ApJ...698.1570L, 2011PhRvE..83a6302L, 2016MNRAS.461.2605G, 2022MNRAS.515L..17Z}. We appeal to a simpler hydrodynamic picture since only two CSO 2s (J0741+2706 and J1326+3154) have observed rotation measures \citep{2007ApJ...661...78G, 2016MNRAS.459..820T}. The polarized regions are concentrated at the heads and have diameters $<10$ pc and magnetic fields $<10$ $\mu$G perpendicular to the propagation direction. Additionally, neither of these sources is classified as a CSO 2.0 \citep{PaperIII}. While obscuring tori may surround and decohere the large-scale polarized emission, low polarization in these examples make a purely hydrodynamic model adequate at this stage. 
 
\subsubsection{Ballistic Projectile Model}
An alternative possible model for early life CSO 2.0s posits that rather than being active jets, the hot spots represent energetic projectiles ejected from the core by a powerful fueling event. In this picture as we envision it, the projectiles, packets of energized plasma injected into the external medium, are produced rapidly by a short-lived jet of negligible duration compared to the CSO lifetime and propagate at near relativistic by their own inertia. We suppose that the projectiles are subsequently confined and decelerated by the external ram pressure as first considered by \cite{1967Natur.216..129D} in the context of Cygnus A. While this model is inadequate for FR-IIs with visible jets, it may apply to some CSO 2.0s which lack obviously visible jets and have small sizes. The acceleration is $\dot{v}_p\approx -\rho_a v_p^2 A_p/M_p$, where $v_p$ is the upward velocity, $A_p$ is the cross-sectional area, and $M_p$ is the mass. The time over which the projectile decelerates to the ambient sound speed from $c$ will be
\begin{equation}
\begin{split}
    t\approx&1000 \text{ yr} \\& \times \left(\frac{M_p}{1 \text{ }M_\odot}\right)\left(\frac{r_p}{10 \text{ pc}}\right)^{-2}\left(\frac{\rho_a}{10^{-24} \text{ g cm}^{-3}}\right)^{-1}\left(\frac{c_s}{250\text{ km/s}}\right)^{-1},     
\end{split}
\end{equation}
where $r_p$ is the radius of the projectile and $c_s$ is the ambient sound speed, while the height above the core will be
\begin{equation}
\begin{split}
    \zeta\approx0.2 \text{ pc}  \left(\frac{M_p}{1 \text{ }M_\odot}\right)\left(\frac{r_p}{10 \text{ pc}}\right)^{-2}\left(\frac{\rho_a}{10^{-24} \text{ g cm}^{-3}}\right)^{-1}\ln \left(\frac{c}{c_s}\right).    
\end{split}
\end{equation}
To get properties commensurate with CSOs, small values of $r_p$ and $\rho_a$ while high values of $c_s$ are needed.
If $\rho_a(\zeta)$ falls off steeply or the projectile area decreases, the deceleration rate slows, allowing the projectile to propagate further into the medium. The lifetime as a CSO 2.0 would depend on when the projectile experiences sufficient adiabatic losses in electron energy. 

\subsection{Mid-life}
\label{sec:midlife}
By mid-life, CSO 2s must transition from nearly relativistic to transonic or subsonic, bridging the CSO 2.0 through CSO 2.2 evolutionary paradigm. The hot spot separation speed of CSO 2.0s ($\sim0.3c$) corresponds to $v_h\approx0.15c$, so the sonic Mach number is $v_h/c_{s}\approx200$ with an ambient sound speed of $c_{s}\approx250$ km s$^{-1}$. 
To respect the observed CSO age and size cutoffs \citep{PaperII}, the Mach number must decrease to order unity in time $t_1< 10^4$ yr without the size increasing by a factor of more than 5. 
While the sonic Mach number in Sedov-like expansion models \citep[e.g.][]{1960SPhD....5...46K, 1998A&A...335..370E, 2009A&A...506.1215O} decreases too slowly, the entrainment of thermal material and the eventual development of turbulence through interface instabilities \citep{1994ApJ...422..542B} can slow down jets sufficiently. Mass loading, as has been proposed to explain the FR-I/FR-II dichotomy in kpc-scale jets \citep{1994MNRAS.269..394K, 1994ApJ...422..542B, 1996ApJ...460..612R, 2002MNRAS.336.1161L, 2014MNRAS.441.1488P, 2020MNRAS.494L..22P, 2021MNRAS.500.1512A}, may therefore partially account for the transition from CSO 2.0 to CSO 2.2. 

With an initial sonic Mach number $M_0\approx200$ and for deceleration over $1000$ yr, momentum conservation implies a mass-loading rate of $0.2$ M$_\odot$ yr$^{-1}$. Assuming a mass entrainment rate $\dot{M}_{a}\approx2\pi\rho_a r\zeta_j c_s$ from the external medium alone, $\rho_a\gtrsim8\times10^{-25}$ g cm$^{-3}$ is  sufficient to slow the expansion, compatible with typical densities in the centers of elliptical galaxies \citep{2010MNRAS.407.1148C} and of our CSO 2.0 environments. Conversely, taking main sequence B star mass-loss rates of $\dot{M}_{\star}\approx10^{-12}-10^{-9}$ M$_{\odot}$ yr$^{-1}$ \citep{2014A&A...564A..70K} as a fiducial value, the number of stars immersed in the lobes needed to provide sufficient mass loading is $\sim10^{11}-10^8$, not exceptionally different from the number within the central kpc of the Milky way galaxy  
\citep{2018A&A...609A..27S}.

Mass-loading could be initiated by various instabilities \citep{2013ApJ...772L...1M, 2017MNRAS.472.1421M, 2023MNRAS.520.3009A, 2023MNRAS.519.1872W}. The lobes naturally expand into an interstellar medium (ISM) with fluctuating density, especially if it is thermally unstable.  Density may vary by 
multiple orders of magnitude on pc scales due to clumping \citep{2014ApJ...789..153L, 2022PhR...973....1D}. The 
instabilities or density variations may cause shocks and turbulent behavior. For stellar mass-loading, numerical simulations have shown that comet-like tails form behind the star where the stellar wind mixes with the jet \citep{2023arXiv230605864P}. Particle acceleration at the mixing sites \citep{1997MNRAS.287L...9B, 2019A&A...623A..91T} or jet-induced disruption of the immersed stars \citep{2012A&A...539A..69B} can cause gamma-ray flares, so high energy non-thermal transients from CSO 2s can corroborate this model. 
The high surface brightness regions in the middle of CSO 2.2 J0131+5545 \citep{2020ApJ...899..141L, PaperIII} may represent the sites of an instability \citep{1991MNRAS.250..581F, 1991ApJ...372..646C, 1993ApJ...405L..13D} or possibly stellar mass-loading.

\subsubsection{Discussion of CSO 2.1s}
CSO 2.1s  resemble both CSO 2.0s and CSO 2.2s and have been classified by \cite{PaperIII} as "mid-life". The catch-all nature of the CSO 2.1 class may include sources that are not necessarily mid-life. CSO 2.1s may develop in asymmetrical environments with different density gradients or nonzero ambient medium velocity, which can induce turbulent mixing in one jet but not the other. Bearing in mind our currently limited understanding, our model will simply assume that the speed at the end of the CSO 2.0 phase decelerates at a constant rate to the sound external speed $c_s$. With deceleration occurring from time $t_1$ to $t_2$, the distance to the working surface from the central engine is
\begin{equation}
\label{eq:cso21size}
\begin{split}
    \zeta(t)=\frac{c_s-v_h(t_1)}{t_2-t_1}&\left(\frac{t^2-t_1^2}{2}-t_2\left(t-t_1\right)\right)\\&+v_h(t_1)(t-t_1)+\zeta(t_1).  
\end{split}
\end{equation}
We assume that the jet power is negligible at this stage, so the lobe energy should be approximately the total energy of the fueling event.

\subsubsection{Jet Inclination Bias in the CSO 2.1 Population}
Because CSO 2.1s exhibit features common to both CSO 2.0s and 2.2s, some true CSO 2.2s may be classified as CSO 2.1s due to the inclination of the radio jets on the plane of the sky. If the jet axis is inclined to the sky plane, one CSO lobe will be closer to Earth than the other, producing a small difference in the radio emission travel time. While the CSO selection criteria make it unlikely that CSO 2s are highly inclined to the sky plane, CSO inclinations may reach as high as $45^\circ$ and fit selection criteria. The delay in light travel time between the two sides should be $\Delta \tau=D\sin i/c$, where $i$ is the angle between the jet axis and the plane of the sky. The sky projected source size should be $D_{proj}=D \cos i$. In terms of projected size, the time delay between the emission from the two lobes is $\Delta\tau=D_{proj}\tan i/c$. With maximum $i=45^\circ$ and $D_{proj}\sim600$ pc, an upper limit on the delay time is $\Delta \tau\sim2000$ yr. Assuming CSO 2s live for $10^4$ yr and spend $2/5$ of their lifetime as CSO 2.0s and CSO 2.2s and $1/5$ as CSO 2.1s, at most half of true CSO 2.2s would be misclassified as CSO 2.1s due to the time delay. In the observed distribution of 43 \textit{bona fide} CSO 2s with spectroscopic redshifts, CSO 2.1s make up a larger fraction of the CSO 2 population than CSO 2.2s, suggesting that this bias may have some effect. For example, if CSO 2.0s and 2.2s each intrinsically account for 40\% of the CSO population, 35\% of CSO 2.2s have been misclassified as CSO 2.1s. This would suggest an average time delay of 1400 yr across the population. The bias cannot affect the entire population because CSO 2.1s make up more than $20\%$ (the maximum percentage of the source lifetime affected by viewing inclination) of the observed CSO population. Therefore,  a true medium class of CSO 2s should exist. 
\subsection{Late Life}
\label{sec:latelife}
The apparent motion of CSO 2.2s is invisible, so their upward expansion is likely transonic or subsonic. We suppose that CSO 2.2s correspond to convectively rising turbulent plumes. In a convectively stable ambient medium, the plume will reach a terminal height where the ambient and plume densities equilibrate. The plume then spreads horizontally rather than vertically \citep{1956RSPSA.234....1M}, producing the mushroom cloud shape characteristic of volcanic and other geophysical plumes \citep{2010AnRFM..42..391W}. With a decaying ambient density profile  (e.g. $\rho_a=\rho_0\left(\zeta_0/\zeta\right)^a$),  this behavior should be generic. In fact, M87 displays similar features $\sim20$ kpc from its core \citep{2000ApJ...543..611O}. In many CSO 2s, the equilibrium height could be much closer to the core, even within the central kpc after mass-loading. The lateral expansion of CSO 2.2s signals that they may have already reached their equilibrium height.

Similar to plumes, some AGN show evidence of previous jet activity in the form of coherent bubbles \citep{2000ApJ...543..611O, 2001ApJ...554..261C, 2002ApJ...568..163N, 2003MNRAS.344L..43F}, which rise at their terminal velocity \citep{2001ApJ...554..261C, 2019MNRAS.488.4926I}
\begin{equation}
    v_t=\sqrt{2 g \frac{V}{C_d A}\frac{(\rho_a-\rho_b)} {\rho_a}}\approx c_s,
\end{equation} where $g$ is gravitational acceleration, $V$ is the bubble volume, $C_d$ is the drag coefficient, $A$ is the bubble cross-sectional area, $\rho_a$ remains the ambient density, and $\rho_b$ is the density inside the bubble. Some CSO 2.2 images morphologically resemble simulations of rising bubbles \citep{2001ApJ...554..261C, 2002MNRAS.332..271R, 2008MNRAS.389L..13S, 2023MNRAS.521.4375H}. Although difficult to reconcile with mass-loading-induced turbulence, bubbles may form, perhaps with the assistance magnetic draping \citep{2008ApJ...677..993D, 2018ApJ...857...84B}, and spread laterally. This would dilute the energy and explain the lack of large low power CSO 2.2s in the \cite{PaperIII} population.

\subsubsection{Subsonic Rise and Disappearance of CSO 2.2s}
\label{subsubsec:subsonic}
A coarse model of subsonic CSO 2.2s has the lobes expand constantly at the ambient sound speed. The duration of the CSO 2.2 phase must be comparable to that of the other CSO phases to respect the population distribution observed by \cite{PaperIII}. For reasons that remain unclear, CSO 2s must disappear at a certain point. One possibility, which we consider and elaborate on further in Sec. \ref{sec:Disappearance}, is the existence of a steep density gradient at some characteristic point in the host galaxies. The pressure of the ambient medium and by extension the pressure inside the CSO will drop considerably, causing the lobes to expand laterally and quickly form the mushroom cloud shapes discussed previously. This would cause CSO 2s to rapidly fade and partially facilitate the observed CSO size cutoff. The pressure surrounding a CSO may drop by a factor $n$ when reaching this stage, so that $P_{ 2.2}=P(t_2)/n$ or alternatively, $V_{2.2}=n^{3/4}V(t_2)$ (for an internal adiabatic index of 4/3). While the height may not change appreciably, $n^{3/4}$ accounts for the factor by which the cross sectional area of the lobes increases above the pressure drop. This idea is separate from the jet fueling scenarios we consider and not mutually exclusive with those scenarios.

\subsection{Numerical Realizations of the Model}
We present the results of the toy model for CSO 2s we have outlined above. The injection of a finite energy amount $E_0$ in a finite time gives rise to the jet power profile in Sec. \ref{subsubsec:ActiveJet}. The CSO 2s produce synchrotron radio emission with equipartition radio luminosity 
\begin{equation}
\label{eq:luminosity}
    L(t)= C(\alpha) V(t)^{-3/4} E(t)^{7/4},
\end{equation}
where $C$ is a numerical coefficient which is a function of the spectral index $\alpha$, while $V(t)$ is the emitting volume, and $E(t)$ is the energy in the CSO at time $t$.

We assume that CSO 2s evolve through the CSO 2.0--2.1--2.2 paradigm. In our realization, we apply the active jet model for CSO 2.0s, lasting from $t=0$ to $t=t_1$, with $t_0<t_1$. The source then becomes an inactive CSO 2.1 during $t_1<t<t_2$ and a CSO 2.2 when $t>t_2$. We compute the energy and volume of CSO 2.0s in the manner described in Sec. \ref{subsubsec:ActiveJet}. The volume of CSO 2.1s is $V\propto \zeta^3$ with $\zeta$ computed by eq.~\ref{eq:cso21size}. For CSO 2.2s, we implement the two different cases discussed in Sec. \ref{subsubsec:subsonic}: 1) the volume of the CSO evolves as $V\propto \zeta^3$ and 2) when CSO 2.1s transition to CSO 2.2s, the pressure drops by a factor $n$, corresponding to an increase in cross-sectional area by a factor of $n^{3/4}$ without an increase in $\zeta(t)$. In both cases, we set $\zeta(t)=c_s t+\zeta(t_2)$.

Because the synchrotron cooling times are comparable to the source lifetimes, we correct the energy available for CSO 2.1s and 2.2s. Assuming $L=-\frac{dE}{dt}$, we integrate eq.~\ref{eq:luminosity} from $t_1$ to $t$, keeping the volume fixed at $V(t_1)$ (since CSO 2s grow more slowly in these phases), and obtain the energy in the inactive jet phases
\begin{equation}
\label{eq;correctedenergy}
    E(t)=E_0 \left(1+\frac{3}{4}\frac{L(t_1)}{ E_0}(t-t_1)\right)^{-\frac{4}{3}},
\end{equation}
where $L(t_1)$ is the value of eq.~\ref{eq:luminosity} evaluated at $t_1$. This is the energy we substitute into eq.~\ref{eq:luminosity} and is less than $E_{0}$.

We choose a set of parameters to qualitatively reproduce the observed distribution of CSO 2s on the $(P, D)$ plane as shown in Fig.~1 of \cite{PaperIII}. We draw the timescales from uniform distributions with $t_0\in\{2000 \text{ yr}, 3000 \text{ yr}\}$, $t_1\in\{t_0+1000\text{ yr}, t_0+3000 \text{ yr}\}$, and $t_2\in\{t_1+2000 \text{ yr}, t_1+4000 \text{ yr}\}$. The ambient density profiles are set by uniformly drawing $\rho_a\in\{10^{-23}  \text{ g cm}^{-3}, 10^{-22} \text{ g cm}^{-3}\}$ with $\zeta_0=5$ pc and $a=1$.  We fix $p_1=4/3$ and $p_2=5/3$. We fix the spectral index $\alpha=0.7$ and calculate the luminosity at each simulated source's synchrotron self-absorption frequency. We set the ambient sound speed as $c_s=250$ km/s, which corresponds to an ISM temperature $\sim10^7$ K. To generate a population of CSO 2s, we draw $E_0\in\{0, 10 M_\odot c^2\}$, $\theta_0\in\{10^\circ, 20^\circ\}$, and $t\in\{0, 20000 \text{ yr}\}$, uniformly. For the model in which CSO 2.2s experience a steep pressure drop, we uniformly draw $n\in\{0, 100\}$.

\begin{figure*}
    \centering
    \includegraphics[width=1.0\linewidth]{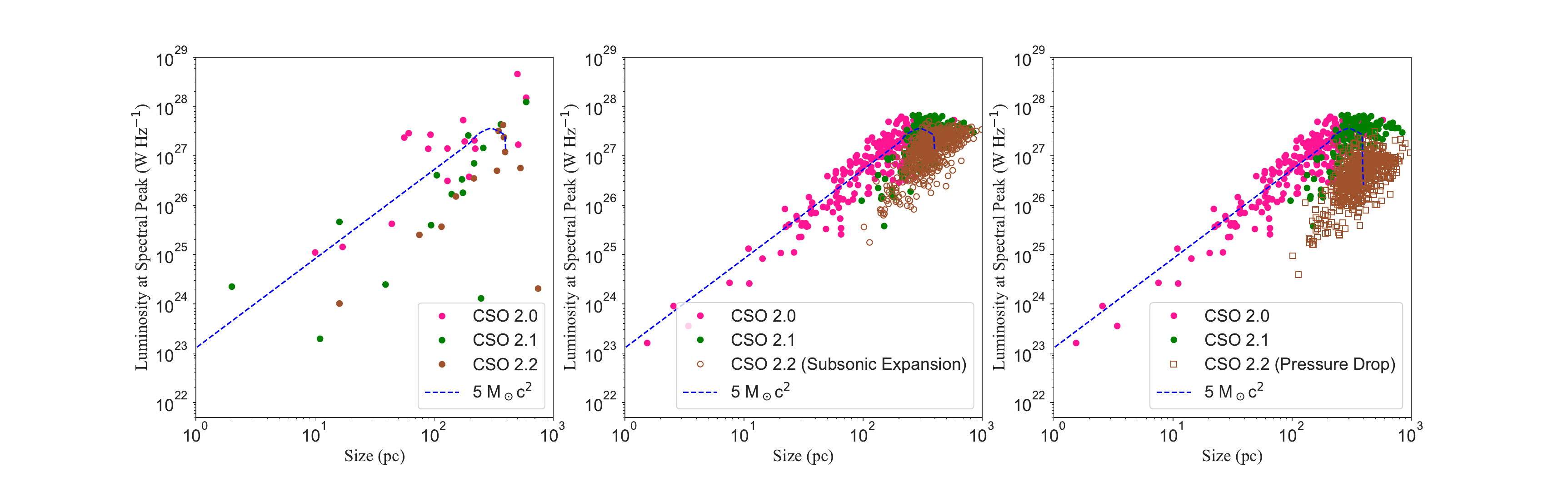}
    \caption{The $(P, D)$ distribution of the real \textit{bona fide} CSO 2 population \citep{PaperIII} (left), our 4-phase toy model in which CSO 2.2 lobes expand at the ambient sound speed (middle), and our model in which a large drop in external pressure causes a proportionate increase in cross-sectional area of CSO 2.2s (right). The dashed blue curve in the each panel show the evolution of a model CSO with $E_0=5$ $M_\odot c^2$, $\theta_0=15^\circ$, and all environmental properties set by their median values.}
    \label{plt:ModelPD}
\end{figure*}

\begin{figure*}
    \centering
    \includegraphics[width=1.0\linewidth]{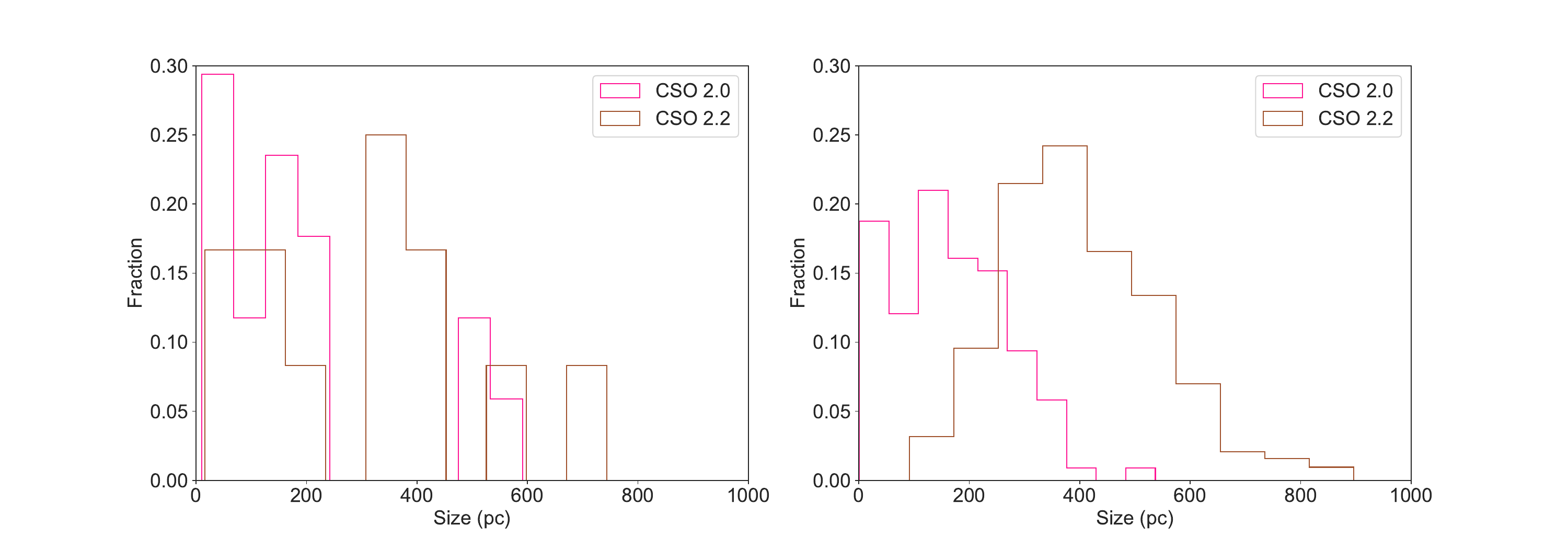}
    \caption{The size distributions of the \textit{bona fide} CSO 2.0 and 2.2 populations \citep{PaperIII} (left), and the size distributions of our simulated CSO 2.0 and 2.2 populations (right).}
    \label{plt:size}
\end{figure*}

The populations of 1000 simulated CSO 2s along with the 43 \textit{bona fide} CSO 2s with spectroscopic redshifts are shown on the $(P, D)$ plane in Fig. \ref{plt:ModelPD}. 
Like the real population, there is a divide between CSO 2.0s which occupy the upper-left portion of phase space, while CSO 2.2s occupy the lower-right portion. We show the size distributions of our model CSO 2.0s and 2.2s in Fig. \ref{plt:size}. Our model choices, particularly the maximum energy, recreate the size distribution cutoff. Our choice of $a=1$, consistent with the observations of CSO 2s, can also be approximately derived by assuming that the radiative cooling time of the ambient gas is proportional to its infall time from height $\zeta$ at fixed temperature. This assumption is broadly consistent with observations of gaseous media in the centers of galaxies \citep{2022PhR...973....1D}.

By design, the two proposed models have a different spread of CSO 2.2s across the $(P,D)$ plane.
The middle panel of Fig. \ref{plt:ModelPD} shows a buildup of CSO 2.2s very near the CSO 2.1 population. In contrast, the model with a pressure drop shows a broader spread of CSO 2.2s, placing some older CSO 2s firmly below the flux limits of the surveys from which the real CSO population is gathered. The model realizations might suggest host galaxy features at radii $\sim100-1000$ pc which cause rapid lateral expansion and a commensurate drop in luminosity. We elaborate on the hypotheses in Sec. \ref{sec:Disappearance}.

\section{The Disappearance of Compact Symmetric Objects}
\label{sec:Disappearance}
Our model in the center panel  of Fig. \ref{plt:ModelPD} shows that CSO 2.2s should fade slowly if embedded in an ambient medium with a continuous density gradient. Evolving our population to times longer than $2\times10^4$ yr gives an even larger buildup of CSO 2.2s and causes CSO 2.2s to vastly outnumber CSO 2.0s, incommensurate with the findings of \cite{PaperIII}. To respect the observed population distribution, CSO 2s should fade rapidly without much height increase. As the right panel of Fig. \ref{plt:ModelPD} shows, a sharp drop in ambient pressure around when CSO 2s transition to CSO 2.2s can produce a broad spread in the CSO 2.2 population, with particularly old CSO 2.2s ($>10^4$ yr) disappearing below the survey flux limits \citep{PaperI}. Such a rapid decrease in pressure, if real in some sources, may result from some unique host galaxy feature. We propose two possible features which could cause such a drop: 1) the rapid falloff in the stellar density around the galaxy core radius and 2) the presence of shocks separating the inner core from the exterior.
\subsection{Stellar Distribution}
Probes of the stellar distribution may be an excellent way to constrain the ambient environment of CSO 2s. Near and beyond the Bondi radius, the stars are sources of gravitational potential, magnetic fields, and additional outflowing winds \citep{1991MNRAS.248P..14S, 1996ARA&A..34..645M}. Since the stellar distribution in dense environments directly correlates with that of molecular gas \citep[e.g.]{2017A&A...607A..86R}, the stellar distribution broadly characterizes the ambient density. CSO galaxies may possess prominent cores with radii $<1$ kpc, outside of which the stellar density drops substantially.  
Knowledge of core Plummer radii, perhaps obtained from the mass to light ratios in the interiors of CSO galaxies, would strongly constrain this model.

\subsection{Shock Fronts}
Alternatively, density and pressure jumps on account of the buildup of gas at accretion shocks around the cores may affect CSO 2s. In this scenario, infalling gas from galactic halos can cause the formation of shock fronts \citep[e.g.]{2011MNRAS.414.2458V,2023MNRAS.522.5500C} separating the ISM and the core. If these shocks manifest within the central kpc of CSO galaxies, CSO cocoons could blow up upon crossing the density discontinuity. Excess filaments of H$_\alpha$ emission in UV surveys \citep{2014A&A...570A..82V} or IR emission \citep{2008ApJ...685..160N} could point to the buildup of dust in the central kpc of the CSO galaxies and represent tracers of core accretion shocks, should they exist.

\section{Possible Driving Mechanism}
\label{sec:fueling}
\subsection{Tidal Disruption Events} 

\cite{PaperIII} have proposed that stellar tidal disruption events (TDEs) could fuel CSO 2s. The small sizes,  finite lifetimes, and relatively low energies of CSO 2s suggest single catalyzing events as their sources, making TDEs, well established jet launching events  \citep{2011MNRAS.416.2102G, 2014MNRAS.445.3919K, 2019MNRAS.483..565C, 2020NewAR..8901538D}, a plausible candidate. The synchrotron minimum energies $\gtrsim1$ M$_\odot c^2$ in CSO jets are comparable to what a disrupted star can provide. The very large SMBHs at the centers of CSO 2s have the ability to disrupt a variety of stellar populations \citep{PaperIII}; however, the estimated event rates of TDEs and CSO 2s are not quite commensurate. We explore the time scale, fueling, and event rate of CSO 2s in the following subsections.  

\subsubsection{Time Scale}
The fallback time of the debris onto the central SMBH sets the timescale of a TDE \citep{1989IAUS..136..543P, 2009MNRAS.400.2070S, 2011MNRAS.416.2102G, 2012MNRAS.420.3528M}. The disruption is initiated when the star falls within the tidal radius $r_t\approx R_\star (M_{BH}/M_\star)^{1/3}$. The disruption must happen outside the SMBH event horizon, so we may write $r_t$ as
\begin{equation}
    \frac{r_t}{r_g}=2.2  \left(\frac{M_{BH}}{10^8 \text{ }M_\odot}\right)^{-\frac{2}{3}}\left(\frac{M_{\star}}{1 \text{ }M_\odot}\right)^{-\frac{1}{3}} \left(\frac{R_\star}{1 \text{ } R_\odot}\right),
\end{equation}
where $r_g\equiv GM_{BH}/c^2$. The stellar encounter periapse radius $r_p$ must satisfy $r_p<r_t$. The change in specific  energy of the stellar debris is $\Delta e\approx G M_{BH} R_\star/r_p^2\sim G M_{BH}^{1/3} M_\star^{2/3}/ R_\star$ for $r_p\sim r_t$. Therefore, the bound debris will have apoapse distance $r_a\gtrsim r_\star \left(M_{BH}/M_\star\right)^{2/3}$. The characteristic fallback timescale for the debris is \citep{1989IAUS..136..543P}
\begin{equation}
\begin{split}
    t_{f}\sim& \left(\frac{r_a^3}{G M_{BH}}\right)^{\frac{1}{2}} \\&\approx 0.5 \text{ yr} \left(\frac{M_{BH}}{ 10^8\text{ }M_\odot}\right)^{\frac{1}{2}} \left(\frac{M_{\star}}{ 1\text{ }M_\odot}\right)^{-1} \left(\frac{R_\star}{1 \text{ } R_\odot}\right)^{\frac{3}{2}}.    
\end{split}
\end{equation}
Before $t_f$, the mass accretion rate increases rapidly, and should follow $\dot{M}\sim t^{-5/3}$ after $t_f$ \citep[e.g.]{1989IAUS..136..543P, 2009MNRAS.400.2070S}. For main sequence stars, $t_f$ can be of order months to years \citep{2020NewAR..8901538D}. Some class of TDEs should have $t_f\gtrsim1000$ yr for the CSO-TDE hypothesis to hold. 
Giant branch stars with $R_\star>100$ R$_\odot$ give values of $t_f\sim1000$ yr for SMBH masses $\gtrsim10^8$ M$_\odot$. For instance, a 1 $M_\odot$ star with radius of 200 $R_\odot$, commensurate with typical red giant parameters \citep{1994MNRAS.270..121H}, tidally disrupted by a SMBH with $M_{BH}=10^{8.5}$ M$_\odot$ will have $t_f\approx3000$ yr.  In practice, a giant is divided into a dense He core and a less dense envelope, each containing about half the original star's mass, so the stripped envelope mass would supply the infalling  gas while the core remains undisturbed. Nevertheless, the rise in accretion rate over $t_f$ and decay $\dot{M}\sim t^{-5/3}$ is consistent with that assumed in our toy model.

\subsubsection{Energy Supply}
\label{sec:EnergySupply}
The TDE energy extraction efficiency may be $\lesssim10\%$ \citep{2020NewAR..8901538D}, naively suggesting that only massive, difficult to disrupt stars $\sim 10-100$ M$_\odot$ have fuel for CSO 2s. 
However, lower mass TDEs can provide magnetic flux to ignite the extraction of the SMBH spin energy  \citep{1977MNRAS.179..433B, 2013MNRAS.436.3741P}. Shortly after the TDE, we expect the magnetic flux to disperse around the disk and thread the black hole
\begin{equation}
    \Phi_{BH, 0}\approx \Phi_{\star} \frac{R_H^2}{r_p^2},
\end{equation}
where $\Phi_{\star}$ is the flux in the captured star, and $R_H$ is the radius of the SMBH horizon. This magnetic flux coupled with the SMBH spin launches a jet with power \citep{1977MNRAS.179..433B, 2013MNRAS.436.3741P}
\begin{equation}
\label{eq:BZpower}
    L_j=\frac{1}{6\pi c}\Omega_{BH}^2 \Phi_{BH}^2,
\end{equation}
where $\Omega_{BH}$ is the spin of the BH. For a maximally spinning BH, $R_H=r_g$ and $\Omega_{BH}=c/2 r_g$. CSO jet powers can reach $L_j\sim 10^{44}$ erg s$^{-1}$ \citep{PaperIII}, requiring magnetic fields at the SMBH to reach $\sim10$ kG. 

A very strong dipolar dynamo action \citep{1955ApJ...122..293P, 1995A&A...298..934R} or the magneto-rotational instability \citep{1991ApJ...376..214B, 2001A&A...374.1035A} could grow the field after being seeded by the TDE.
The difficulty is that the net magnetic flux must be conserved with vertical magnetic flux randomly walking around that supplied by the TDE \citep{2014MNRAS.437.2744T}. A pre-existing "fossil" disk which already has some seed flux could nevertheless be amplified by the TDE \citep{2014MNRAS.437.2744T}, increasing the jet power to make an observable CSO. 

Once more, giant branch stars provide a compelling progenitor. Asymptotic giant branch stars posses surface magnetic fields as high as $10$ G \citep{2002A&A...394..589V, 2005A&A...434.1029V, 2011ApJ...728..149V} while red giants have surface magnetic fields as high as 300 G \citep{2008A&A...491..499A, 2009ARA&A..47..333D}. The cores of red giants may even support magnetic fields exceeding $10^5$ G \citep{2015Sci...350..423F}. For a single incoming giant with average magnetic field $B_\star\gtrsim10^2$ G, stellar radius $R_\star\approx 200$ R$_\odot$, $r_p=100$ $r_g$, and SMBH mass $10^8$ M$_\odot$, the initial TDE would still only provide a seed magnetic field, producing jet power $L_j\sim10^{33}$ erg s$^{-1}$. 

The released gravitational energy of infalling debris provides sufficient energy to grow the magnetic field in the vicinity of the SMBH. The energy needed to build the magnetic field is $B^2R_{in}^3\approx G M_{BH}m_{\star}/R_{in}$, where $R_{in}$ is the distance from the SMBH to which the TDE material infalls from infinity. A 1 $M_\odot$ debris stream can provide the energy for $B\approx10$ kG by falling to $R_{in}\sim40$ $r_g$. Since a giant branch star can be disrupted at $r_t>100$ $r_g$, the infalling material can release the energy needed to build up the field. Even a $\sim1$ $M_\odot$ star can generate sufficient magnetic field to power CSO jets. At least energetically, a wide range of TDE masses can produce CSO 2s, not simply the most massive. Via a similar argument, the infalling debris of lower mass stars should also be able to provide sufficient pressure to hem in the magnetic field at the SMBH.

\subsubsection{Ballistic Model and TDEs}
The scenario we have just discussed changes slightly under the ballistic projectile model for early CSO fueling. In this case, a TDE with $t_f<<1000$ yr, like that from a main sequence star, would provide the energy injection into the projectile. To inject a full $M_\odot c^2$ into the projectile within this short duration, $L_j$ would need to be at least 2 orders of magnitude larger than assumed in the previous section, requiring magnetic fluxes at the central SMBH to be higher by over a factor of 10. This would consequently require a very efficient mechanism for increasing $\Phi_{BH}$ such as a stronger seed field \citep{2014MNRAS.437.2744T, 2019MNRAS.483..565C}.  

\subsubsection{Event Rate}
The inferred CSO 2 birth rate varies strongly with luminosity, rising from  $\sim 3 \times 10^{-7}$ Gpc$^{-3}$ yr$^{-1}$ at $P = 2\times 10^{28}$ W/Hz to $\sim 2 \times 10^{-3}$ Gpc$^{-3}$ yr$^{-1}$ at $P = 3\times 10^{25}$ W/Hz \citep{PaperIII}. 
The birth rate of jetted-TDEs has been weakly constrained to be (0.003–1)$\times$ the total TDE rate \citep{2020NewAR..8901538D}, which implies a birth rate of $3-10^3$ Gpc$^{-3}$ yr$^{-1}$.  
The TDE sources which can become CSO 2s likely require long duration accretion events as well as preferential jet launching conditions. Particularly, high jet-powering efficiencies (already a requirement, as we have discussed in Sec. \ref{sec:EnergySupply}) and aligned disk-jet angular momenta \citep{2023ApJ...957L...9T} are needed,
suggesting that only a small population of stars have favorable initial conditions. The disproportionate paucity of red giants in galactic centers \citep{1996ApJ...472..153G, 2009MNRAS.393.1016D} suggests an even lower rate for these favorable source candidates. TDEs with long enough timescales also require massive SMBHs like those inferred in CSO galaxies \citep{PaperIII}. Moreover, the flux limits on CSO 2 surveys imply that a larger population of low luminosity sources might exist. Deeper radio surveys will better probe the CSO event rate and could provide results commensurate with the TDE rates.

\subsection{The Unlikelihood of Gas Cloud Captures}
Generically, the infall of cold gas from the host galaxies can feed an AGN \citep{2012ApJ...746...94G, 2021agnf.book.....C}. While an infalling gas cloud with mass $\sim100$ $M_\odot$ destabilized from kpc distances provides both the energy and material needed to power a jet, tidal forces from the SMBH should disrupt the cloud before reaching the central pc. A hot gas cloud at the Jeans limit has tidal radius 
\begin{equation}
    r_t\approx10 \text { pc} \left(\frac{M_{BH}}{10^8 \text{ }M_\odot}\right)^{\frac{1}{3}} \left(\frac{m_{c}}{100 \text{ }M_\odot}\right)^{\frac{2}{3}}\left(\frac{T}{300 \text{  K}}\right)^{-1}.
\end{equation}
Clouds with masses $\sim100$ $M_\odot$  will consequently be tidally disrupted well beyond a few 100 $r_g$. The dynamical time $t_d\approx r_t^{3/2}/G^{1/2} M_{BH}^{1/2}$ with $M_{BH}=10^8$ $M_\odot$ is $\gtrsim10^5$ yr, an order of magnitude too large to explain CSO 2s. In principle, smaller ($m_c\sim10$ $M_\odot$) and hotter ($T\sim3000$ K) gas clouds could have more amenable timescales, although such cases are very similar to single stellar captures. 
\subsection{Testing the Hypothesis}
The sharp size cutoff found by \citet{PaperII} needs to be verified on a larger complete sample of CSO 2s. Work is underway to more than triple the sample size to test this result. Precision measurements of this cutoff as well as multiwavelength observations of the central 100 pc can test the hypothesis.
Recalling the characteristic working surface height $\zeta_c$ from Sec. \ref{subsubsec:ActiveJet} and assuming the source size doubles over the course of its lifetime (as in both the real and our model CSO populations), the maximum diameter of a typical CSO 2 is
\begin{equation}
\begin{split}
   D_c=400 &\text { pc} \left(\frac{E_{tot}}{10 \text{ }M_\odot c^2}\right)^{\frac{1}{4}} \left(\frac{t_0}{2500 \text{  yr}}\right)^{\frac{1}{2}}\\&\times\left(\frac{\theta_0}{15^\circ}\right)^{-1}\left(\frac{\rho_a(\zeta=5\text{ pc})}{5 \times 10^{-23} \text{g cm}^-3}\right)^{-\frac{1}{4}},    
\end{split}
\end{equation}
where we have assumed the density profile has $a=1$. For an initial mass function $\propto m^{-2}$ over 1-100 $M_\odot$ \citep{2001MNRAS.322..231K, 2002Sci...295...82K, 2003PASP..115..763C}, stars with $M_\star<10$ $M_\odot$ should outnumber more massive stars by a factor of $\gtrsim10$. Consequently, assuming roughly similar environmental and jet properties, we predict that CSO 2s with sizes $\lesssim400$ pc will outnumber those with sizes $\gtrsim400$ pc by a factor of $\sim10$ if the TDE hypothesis holds. 

The observation of a population of long duration ($>1$ yr) jetted TDEs (potential proto-CSOs) would directly probe the hypothesis we propose for CSO 2s. A few such events have already been observed \citep{2023arXiv230801965S}, and hopefully soon population studies may be conducted on them.  
Self-collisions of post-TDE debris streams \citep{2021arXiv211203918B, 2022ApJ...931L...6B} may produce additional electromagnetic transients in X-ray, optical, and UV bands \citep{2016ApJ...830..125J,2019arXiv191206081L}. Fading radio afterglow from these collision sites may be observable as localized hot spots in particularly young CSO 2s cores. Additionally, the UV and optical spectra should reflect the elemental abundances deposited by a stellar capture. An over-abundance of nitrogen relative to carbon \citep{2016MNRAS.458..127K, 2023MNRAS.523.3516C}, in particular, would be expected from old progenitor stars and such an observation would strongly support our giant star TDE hypothesis.

\section{Conclusion}
\label{sec:conclusion}
With the current state of the observations, the phenomenology of CSO 2s remains ripe for study.
As a class of jetted-AGN with potentially distinct fueling mechanisms and host environments, CSO 2s represent excellent galactic center probes and reveal the evolution of radio loud AGN. In this work, we have considered the possible mechanisms involved in CSO 2s and presented a semi-analytic model of their life cycle. We have mapped the evolutionary sequence presented by \cite{PaperI, PaperII, PaperIII} onto the phases of the model and explored the phenomenological implications of the different CSO sub-classifications. 

At present, the most pressing questions about CSO 2s concern how sharp the apparent size cutoff is, how they are fueled, and what causes them to fade.  CSO 2s are active jets for some period of time; for how long, however, remains an open question. Eventually, CSO 2s with finite fuel should decelerate to subsonic speeds as their power source is extinguished. This must occur to prevent CSO 2s from reaching sizes exceeding 1 kpc in total. We have discussed two fueling possibilities: TDEs or the infall of large molecular clouds. Given the current evidence, we favor the TDE interpretation and suggest that TDEs of giant branch stars can naturally explain the timescales and event rates of CSO 2s. It remains possible that much shorter TDEs of main sequence stars can reproduce the morphology. Multiwavelength observations of CSO 2s, particularly in the infrared with {\it James Webb Space Telescope} or in UV, coupled with continued VLBI studies should strongly probe the progenitor models and the properties of the host galaxies. Furthermore, our results motivate pointed hydrodynamic and magnetohydrodynamic jet launching simulations to see if the fueling scenarios we posit reproduce the observed radio morphologies.

More broadly, CSO 2s may represent a model system for studying the lifespans of double radio sources, especially because they probe the critical phase of evolution in which the luminosity increases with time as the jet power increases. As a prevalent class of smaller finitely fueled radio loud AGN, CSO 2s, including those which grow into FR-IIs, display a wide variety of morphologies corresponding to their ages and interactions with ambient media. Observations of CSO 2s at all stages of their lifespan may provide insights into the fate of their larger, longer lived counterparts. The late life evolution of CSO 2s may morphologically correspond to the end state of FR-I or FR-II sources when their central engines expire. The continued study of CSO 2s has the potential to answer questions regarding galactic astrophysics, the life cycle of radio jets, and the dynamics of accretion disk formation. The study of CSO 2s  brings time domain astronomy into the study of relativistic jets in AGN.  Identifying small CSO 2s early, we might follow the evolution of jets in individual systems. CSO 2s open a window for galactic and high energy astrophysical discoveries.

\section*{Acknowledgements}
The authors are grateful to Roger W. Romani for useful discussions and the referee for helpful comments that improved the manuscript. A.S. acknowledges the support of the Stanford University Physics Department Fellowship and the National Science Foundation Graduate Research Fellowship. This work was supported by a grant from the Simons Foundation (00001470, R.B., A.S.). In recognition of his many important contributions to astrophysics and cosmology, and in particular of his work on jetted-AGN, this paper is dedicated to Mark Birkinshaw. Mark was working hard on this paper right up until his death from cancer on 23 July 2023. Mark is sorely missed world-wide by his colleagues and friends.

\section*{Data Availability}
The data underlying this article will be shared on reasonable request to the corresponding author.
\bibliographystyle{mnras}
\bibliography{refs}{}

\end{document}